\documentclass[fleqn,usenatbib]{mnras}
\usepackage{newtxtext,newtxmath}
\usepackage[T1]{fontenc}

\DeclareRobustCommand{\VAN}[3]{#2}
\let\VANthebibliography\thebibliography
\def\thebibliography{\DeclareRobustCommand{\VAN}[3]{##3}\VANthebibliography}

\usepackage{graphicx}	% Including figure files
\usepackage{amsmath}	% Advanced maths commands
\usepackage{amssymb}	% Extra maths symbols
\usepackage{listings}

\usepackage{scalerel}
\usepackage{tikz}
\usetikzlibrary{svg.path}
\definecolor{orcidlogocol}{HTML}{A6CE39}
\tikzset{
  orcidlogo/.pic={
    \fill[orcidlogocol] svg{M256,128c0,70.7-57.3,128-128,128C57.3,256,0,198.7,0,128C0,57.3,57.3,0,128,0C198.7,0,256,57.3,256,128z};
    \fill[white] svg{M86.3,186.2H70.9V79.1h15.4v48.4V186.2z}
                 svg{M108.9,79.1h41.6c39.6,0,57,28.3,57,53.6c0,27.5-21.5,53.6-56.8,53.6h-41.8V79.1z M124.3,172.4h24.5c34.9,0,42.9-26.5,42.9-39.7c0-21.5-13.7-39.7-43.7-39.7h-23.7V172.4z}
                 svg{M88.7,56.8c0,5.5-4.5,10.1-10.1,10.1c-5.6,0-10.1-4.6-10.1-10.1c0-5.6,4.5-10.1,10.1-10.1C84.2,46.7,88.7,51.3,88.7,56.8z};
  }
}
\newcommand\orcidicon[1]{\href{https://orcid.org/#1}{\mbox{\scalerel*{
\begin{tikzpicture}[yscale=-1,transform shape]
\pic{orcidlogo};
\end{tikzpicture}
}{|}}}}

\newcommand{\change}[1]{#1}

\title[Accelerating exoplanet transit modelling using Taylor series]{Going back to basics: \change{accelerating exoplanet transit modelling using Taylor-series expansion of the orbital motion}}

\author[H. Parviainen et al.]{
H. Parviainen$^{1,2}$\thanks{E-mail: hannu@iac.es}\orcidicon{0000-0001-5519-1391},
J. Korth$^{3}$\orcidicon{0000-0002-0076-6239}
\\
$^{1}$Instituto de Astrof\'isica de Canarias (IAC), E-38200 La Laguna, Tenerife, Spain\\
$^{2}$Dept. Astrof\'isica, Universidad de La Laguna (ULL), E-38206 La Laguna, Tenerife, Spain\\
$^{3}$Rheinisches Institut f\"ur Umweltforschung, Abteilung Planetenforschung an der Universit\"at zu K\"oln, Universit\"at zu K\"oln, Aachenerstraße 209, 50931 K\"oln, Germany\\
}

\date{Accepted XXX. Received YYY; in original form ZZZ}

\pubyear{2020}

\begin{document}
\label{firstpage}
\pagerange{\pageref{firstpage}--\pageref{lastpage}}
\maketitle

\begin{abstract}
A significant fraction of an exoplanet transit model evaluation time is spent calculating projected distances
between the planet and its host star. This is a relatively fast operation for a circular orbit, but slower 
for an eccentric one. However, because the planet's position and its time derivatives are constant for any
specific point in orbital phase, the projected distance can be calculated rapidly and accurately in the
vicinity of the transit by expanding the planet's $x$ and $y$ positions in the sky plane into a Taylor series at
mid-transit. Calculating the projected distance for an elliptical orbit using the four first time
derivatives of the position vector (velocity, acceleration, jerk, and snap) is $\sim100$ times faster 
than calculating it using the Newton's method, and also significantly faster than calculating $z$ for 
a circular orbit because the approach does not use numerically expensive trigonometric functions. 
The speed gain in the projected distance calculation leads to 2-25 times faster transit model evaluation 
speed, depending on the transit model complexity and orbital eccentricity.
Calculation of the four position derivatives using numerical
differentiation takes $\sim1\,\mu$s with a modern laptop and needs to be done only once for a given
orbit, and the maximum error the approximation introduces to a transit light curve is below 1~ppm for 
the major part of the physically plausible orbital parameter space. 
\end{abstract}


\section{Introduction}

\change{An exoplanet transit model aims to reproduce the photometric signal caused by a planet crossing over the
limb-darkened disk of its host star \citep{Mandel2002,Seager2003,Winn2010e}.}  Evaluation of the  transit model 
can generally be divided into two parts: a) calculation of the projected planet-star centre distance, $z$; and b) 
calculation of the flux decrement caused by a planet occluding a part of the stellar disk visible to the observer. 

The main focus in transit model development has been on the second part, but the calculation
of projected distances can actually take a significant fraction of the total model evaluation time. The standard
approach for calculating $z$ for a single point in time requires several ($\approx6$) trigonometric function calls
and solving the Kepler's equation numerically. While worrying about the computational cost of using trigonometric 
functions might seem frivolous, $z$ needs to be calculated at least once for each photometric data
point when evaluating an exoplanet transit model, and multiple times if the model needs to be supersampled 
(such as for \textit{Kepler} and \textit{TESS} long cadence light curves, \citealt{Kipping2010c}). 
Further, it is already common to have 
photometric data sets of tens or hundreds of thousands of data points (such as a four-year \textit{Kepler} light curve),
and a transit light curve analysis consisting of a posterior optimisation and Markov Chain Monte Carlo
(MCMC) sampling steps can require the model to be evaluated a large number ($\sim10^6$) of times over all the 
data points.

Thus, while calculating $z$ using the standard approaches is a trivial matter for small data sets, speeding 
up the calculation has a potential to yield significant real-life performance gains when modelling modern data 
sets. While accuracy is more important than speed for a scientific code, a speed increase without any significant
sacrifices in accuracy gives freedom for exploratory analyses and experimentation, which can lead to
new interesting discoveries, or, at least, increase the reliability of our analyses.

In this short paper we show how a very simple change in the computation of $z$ can lead to a significant speed-up 
of a transit model without sacrificing model accuracy.
The approach is based on high-school level mathematics (Taylor series expansion, \change{a tool that has been used 
in astronomy and astrophysics for centuries, especially in the research of eclipsing binaries}) and has been tested
thoroughly.
The approach still requires the ability to calculate the eccentric anomaly to a high precision 
in order to calculate the position derivatives using numerical differentiation, but this needs to be done only for 
\change{a small number of points in time (seven in our implementation) for a single Keplerian orbit, rather than
calculating it for each datapoint separately}. 

We provide an example \textsc{Python} implementation of the method in Appendix~\ref{sec:implementation}, 
and the approach has been adopted as the main $z$ computation method in the
\textsc{PyTransit}\footnote{\url{https://github.com/hpparvi/PyTransit}} \citep{Parviainen2015} transit modelling package.

\section{Theory}

\begin{figure}
	\centering
	\includegraphics[width=\columnwidth, clip, trim=0 0.7cm 0 0.7cm]{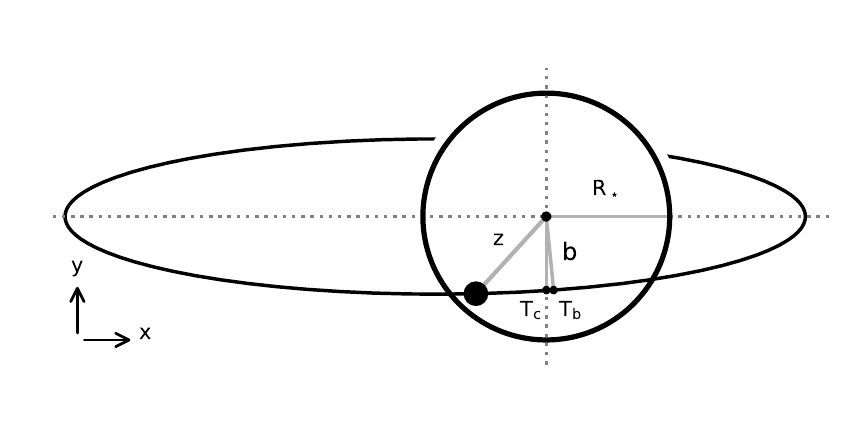}
	\caption{Orbit of a transiting short-period exoplanet on an eccentric orbit. The figure shows the 
	projected planet-star centre distance
	($z$), the impact parameter ($b$), the stellar radius ($R_\star$), transit centre time ($T_\mathrm{c}$),
	and the time of minimum projected distance ($T_\mathrm{b}$). The two latter are equal for a circular
	orbit, but generally differ slightly for an eccentric orbit. Vertical dotted line shows $x=0$ and
	horizontal dotted line shows $y=0$.}
	\label{fig:orbit}
\end{figure}

Calculation of the projected planet-star separation ($z$, see Fig.~\ref{fig:orbit}) as a function of time 
is a necessary step for exoplanet transit model evaluation. The standard approach for calculating $z$ for
a generic eccentric orbit requires the calculation of the eccentric anomaly from the mean anomaly, which requires 
us to solve Kepler's equation, for which no closed-form solutions exist. Thus, the Kepler's equation needs to
be solved using numerical methods, such as iteration or the Newton's method. After the eccentric anomaly
has been solved, the computation of $z$ still requires six trigonometric function calls, which are relatively
expensive operations. 

\begin{figure*}
	\centering
	\includegraphics[width=\textwidth]{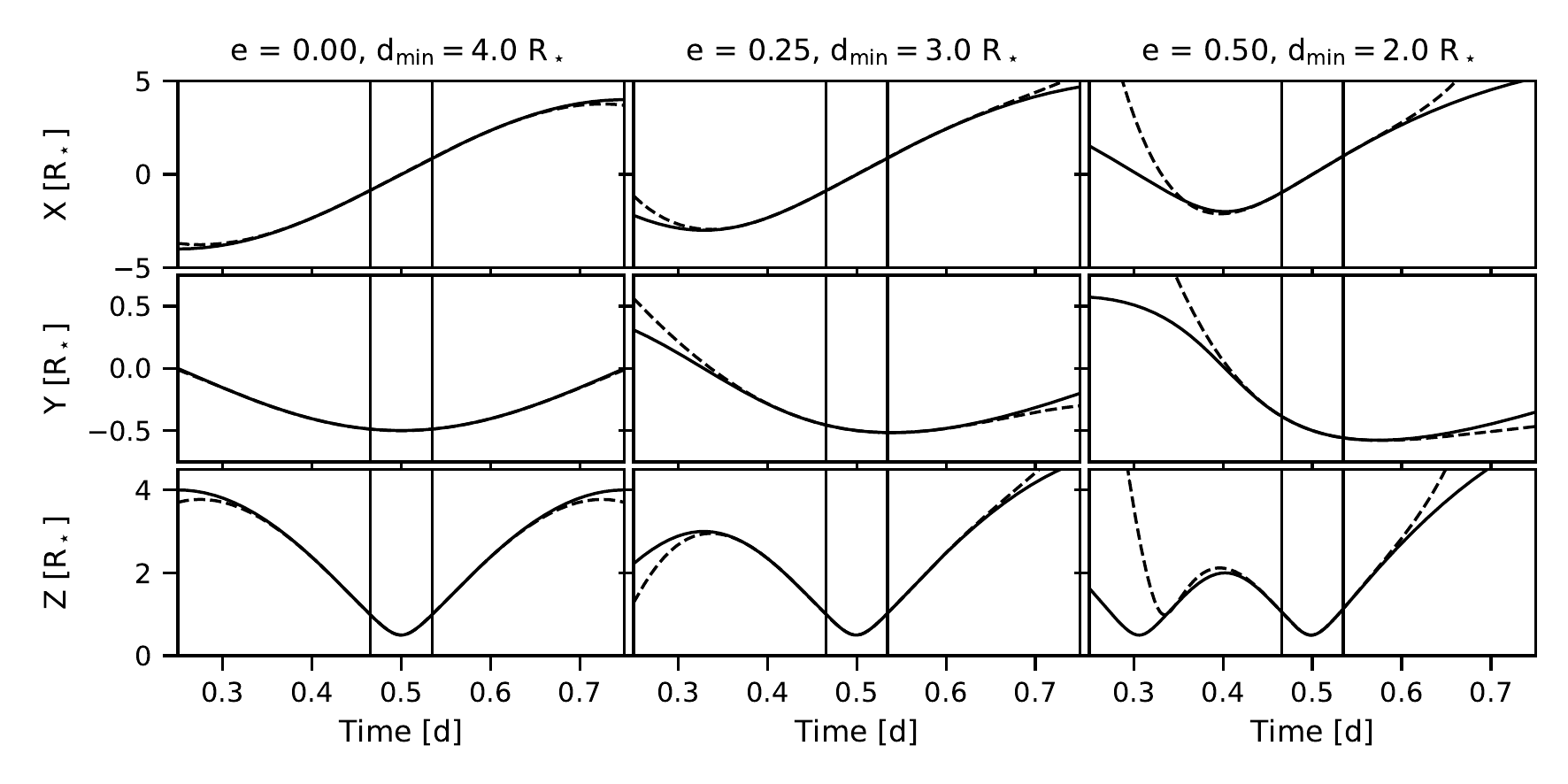}
	\caption{The exact (solid black line) and approximate (dashed black line) sky-plane $x$ and $y$ values and 
	the projected distance $z$ for three short-period orbits with different eccentricities. The vertical lines 
	mark the beginning and the end of a transit. The orbits have a common period (1~d), semi-major axis
	(4~$R_\star$), impact parameter (0.5), and argument of periastron (0). The minimum planet-star separation,
	$d_\mathrm{min}$, tells the separation between the planet and the star at periastron.}
	\label{fig:error1}
\end{figure*}

Could there be a way to calculate $z$ without the need to solve Kepler's equation or use trigonometric
functions? Planet's $x$ and $y$ positions in the sky-plane draw smooth and well-behaved curves
as a function of time, as shown in Fig.~\ref{fig:error1}. The $x$ position is a monotonically increasing
function of time near the transit, and the $y$ position is a smooth unimodal function with a single minimum 
near the transit (this for non-zero impact parameter since $y$ is constant for $b=0$).
These factors mean that the positions can likely be accurately approximated with low-order polynomials near 
the transit.

Thus, we choose to use a Taylor series expansion to represent the planet's position in the sky plane as a 
function of time,
\begin{equation}
    l(t) = \sum_{n=0} \frac{l^{(n)}(t)}{n!} (t - t_0)^n,
\end{equation}
where $l$ is the position (either $x$ or $y$), $t_0$ is the point around which the Taylor series is expanded,
$l^{(n)}$ is the nth derivative of $l$ evaluated at point $t_0$, and $n!$ is the factorial of $n$. 
Mid-transit time where $x=0$ is a natural choice for $t_0$ (although other possibilities exists,
such as the time of minimum projected distance or the time of minimum $y$ position), after which
we only need to select $n$ to ensure a sufficient accuracy so that the approximation does not affect
the transit model in any significant fashion. After testing the accuracy of different $n$ (see discussion
about accuracy later in Sect.~\ref{sec:accuracy}), we chose to use the four first time derivatives of 
position: velocity, acceleration, jerk, and snap.

The first step is to calculate the planet's position in the sky-plane at mid-transit time and its four 
time derivatives. For a circular orbit, the position at mid-transit is $[0, -b]$, where $b$ is the impact
parameter. However, for an eccentric orbit the $y$ position differs from $-b$. 

We use a seven-point central finite difference
method \citep{Fornberg1988} to calculate the 
derivatives. This requires us to
calculate the positions at seven uniformly spaced times centred around the mid-transit time. The calculation
of these locations requires using a standard accurate method for evaluating Keplerian orbits, but this needs
to be done only once for a given orbit.

The velocity, acceleration, jerk, and snap vector component $d$ (either $x$ or $y$) can be computed given 
the points $\vec{l}_i = \vec{l}(t_0 - ih)$ where $i = [-3, -2, -1, 0, 1, 2, 3]$ and $h$ is the time step, 
as\footnote{\change{We group the expressions slightly differently for the actual implementation to reduce 
sensitivity to floating point round-off errors, as shown in Appendix~\ref{sec:implementation}.}}
\begin{align}
    v_d &= \frac{-d_{-3} +  9d_{-2} - 45d_{-1} + 45d_1 - 9d_2 + d_3}{60h},\\
    a_d &= \frac{2d_{-3} - 27d_{-2} + 270d_{-1} - 490d_0 + 270d_1 - 27d_2 + 2d_3}{180h^2},\\
    j_d &= \frac{ d_{-3} -  8d_{-2} + 13d_{-1} - 13d_1 + 8d_2 - d_3}{8h^3},\\
    s_d &= \frac{-d_{-3} + 12d_{-2} - 39d_{-1} + 56d_0 - 39d_1 + 12d_2 - 2d_3}{6h^4}.
\end{align}
After the derivatives have been calculated, the projected distance can be computed for time $t_c$
by first calculating the time difference to the nearest transit centre, $t$,
\begin{align}
    E &= \left \lfloor \frac{t_c - t_0 + 0.5 p}{p}\right\rfloor, \\ 
    t &= t_c - (t_0 + Ep),
\end{align}
where $E$ is the epoch, $t_0$ the mid-transit time, $\lfloor \rfloor$ denotes the floor operation,
and $p$ the orbital period, and then evaluating the Taylor series at $t$ as
\begin{align}
    \vec{l} &= \vec{l}_0 + \vec{v} t + \frac{1}{2} \vec{a} t^2 + \frac{1}{6} \vec{j} t^3  + \frac{1}{24} \vec{s} t^4, \\
    z &= |\vec{l}|
\end{align}
where $\vec{l}_0$ is the position vector at mid-transit, and $\vec{v}$, $\vec{a}$, $\vec{j}$, and  $\vec{s}$
are the velocity, acceleration, jerk, and snap vectors, respectively.

The approximation requires that the planet's position is evaluated at seven points in time for an orbit that
does not evolve in time (that is, the orbital parameters do not evolve in time). However, if the orbit is 
perturbed by external forces, such as other massive bodies in a multiplanet system, the series terms need
to be calculated separately for each transit. \change{This leads to a photodynamical model where the
terms are calculated using a set of positions calculated with an n-body integrator}, as done 
in \textsc{PyTTV} by Korth et al. (2020, in preparation). 

The derivatives known, the computation of $z$ requires only multiplications, summation,
and a single square-root operation. Given the simplicity of the approximation, we provide an example
\textsc{Python} implementation in Appendix~\ref{sec:implementation}.

\section{Performance}
\label{sec:performance}

The real-world improvement in the transit model evaluation speed depends on how heavy the transit shape model is 
relative to the $z$ calculation method (that is, how large fraction of the transit model execution time is spent 
on computing the orbit). For the transit model assuming quadratic stellar limb darkening by \citet{Mandel2002},
the speed gain is between 6 (eccentric orbit calculated using the Newton's method) and 2 (circular orbit), that is,
the model is 6 times faster to evaluate for an eccentric orbit when $z$ is calculated using a Taylor series expansion 
rather than Newton's method. For the most simple transit shape model that assumes uniform stellar disk, 
the speed gain is between 24 (eccentric orbit) and 2 (circular orbit). In both cases, the minimum speed gain is 
around 2 (that is, the model is at least twice as fast to calculate).

\section{Accuracy}
\label{sec:accuracy}

While the Taylor series approximation of $z$ is significantly computationally faster than the other
approaches for calculating $z$, its practical usability depends on the error caused to the exoplanet 
transit model.  The accuracy of the approximation depends on the three-dimensional curvature of the orbit at
the mid transit time, what again depends on the semi-major axis, eccentricity, and argument of periastron.

\begin{figure}
	\centering
	\includegraphics[width=\columnwidth]{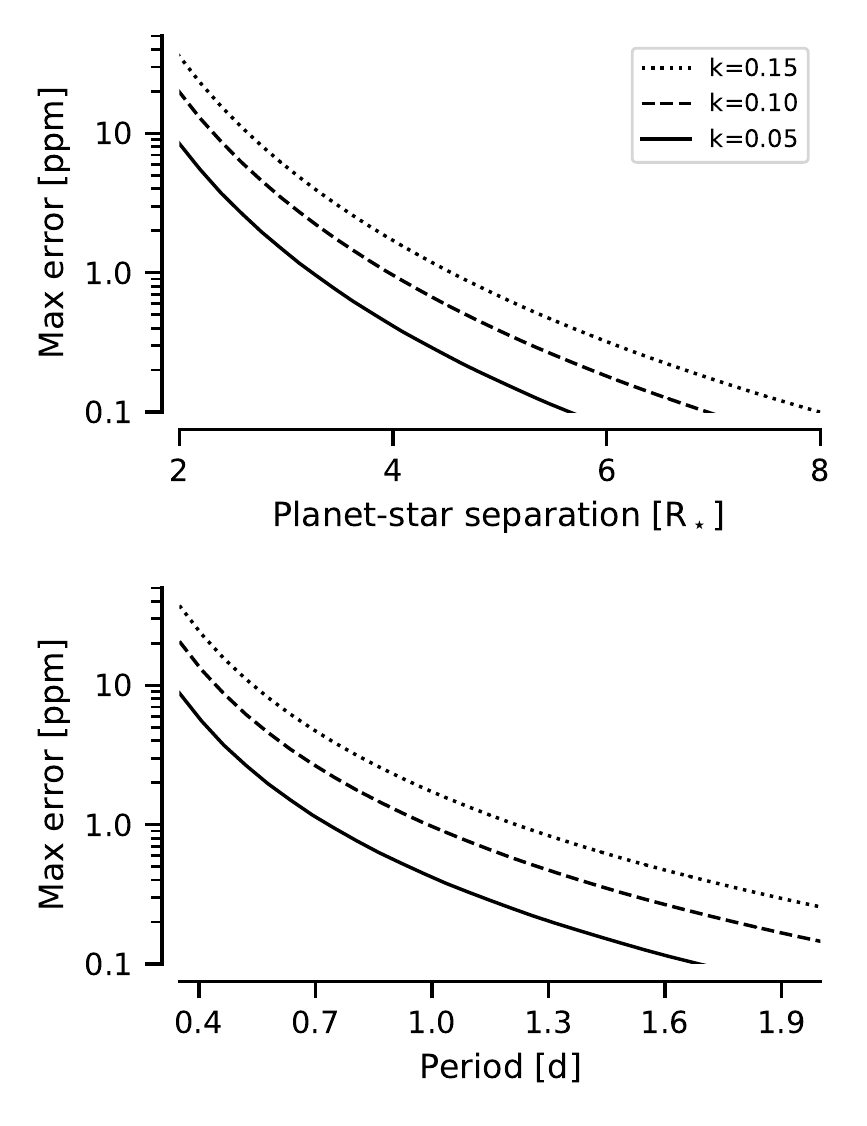}
	\caption{Maximum absolute error to a transit light curve introduced by the approximation 
	         for a circular orbit. The maximum error is shown for three planet sizes as a function
	         of the planet-star separation (upper panel) and period (lower panel) assuming a stellar
	         density of $1.2$~g\,cm$^{-3}$, impact parameter of 0.5, and quadratic limb darkening
	         with coefficients $(u=0.24, v=0.10)$.}
	\label{fig:error2}
\end{figure}

Figure~\ref{fig:error1} shows the actual and approximated $x$ and $y$ coordinates and the projected distance
$z$ for three increasingly eccentric short-period orbits with orbital period, $p$, of 1~d, scaled semi-major axis,
$a$, of $4\,R_\star$, and impact parameter, $b$, of 0.5.
Figure~\ref{fig:error2} shows the maximum absolute errors in a transit light curve caused by the approximation
for a circular orbit as a function of the planet-star separation (that is, the semi-major axis) at mid-transit
(upper panel) and orbital period (lower panel). The orbits correspond to three planets with radius ratio of
0.15, 0.1, and 0.05 orbiting a star with a stellar density, $\rho_\star$, of $1.2$~g\,cm$^{-3}$ with an impact parameter of 0.5. The 
figure focuses on the ultra-short-period and short-period regime because the error is below 1~ppm for
semi-major axes larger than 5~$R_\star$. Considering the currently known transiting exoplanets, the maximum
absolute error introduced by the approximation would be $\sim10$~ppm, staying below 1~ppm for all but the 
most extreme ultra-short-period planets.

\begin{figure}
	\centering
	\includegraphics[width=\columnwidth]{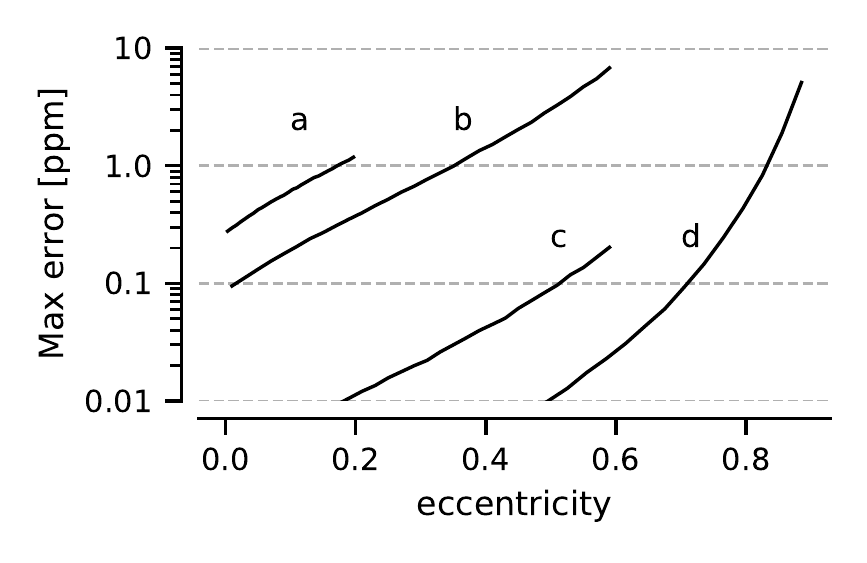}
	\caption{Maximum absolute error to a transit light curve introduced by the approximation 
	for a circular orbit for four different scenarios averaged over $2\times10^5$ samples in 
	argument of periastron and impact parameter.
	The scenarios a, b, c, and d are described in Sect.~\ref{sec:accuracy}}
	\label{fig:error3}
\end{figure}

Figure~\ref{fig:error3} shows the maximum absolute errors in transit light curves caused by the approximation
for four sets of orbital parameters as a function of increasing eccentricity. The scenarios are: a) $p=2.5$~d 
and $a=7.5\,R_\star$, b) $p=5$~d and $a=15\,R_\star$, c) $p=15$~d and $a=25\,R_\star$, and d) $p=30$~d and $a=40.0\,R_\star$, and the
eccentricities cover the range of eccentricities for known planets with periods less or equal to the scenario
period. The maximum error for most of the physically plausible orbits is below 1~ppm, and still below 10~ppm
for the very eccentric orbits.

\section{Conclusions and Discussion}

A planet's normalised planet-star centre distance near a transit (or a secondary eclipse) can be calculated using 
the planet's \change{sky-plane position at mid-transit time} and its four first time derivatives for the whole physically 
plausible orbital parameter space without sacrificing transit model accuracy. The approach is $\sim100$ times 
faster to calculate than an approach using Newton's method to solve the Kepler's equation, and yields a 2-24 gain 
in transit model evaluation speed. \change{Further, since the approach is based on expanding the sky-plane position,
the position can be used directly with transit models that break the radial symmetry, such as the gravity-darkened
transit model for rapidly rotating stars by \citet{Barnes2009}. A gravity-darkened model utilising the approach to
compute the $(x,y)$ position has been added to a coming \textsc{PyTransit} version (v2.4), but here the speed gain
over the standard approach is relatively small due to computational cost of the transit model itself.}

\change{As clear from Figs.~\ref{fig:error2}~and~\ref{fig:error3}, the errors introduced by the approximation into the
transit model are negligible.} The absolute maximum error is below 1~ppm for all but the shortest orbital periods 
and highest eccentricities, and generally below 10~ppm for any currently known planets.

\change{We could also expand $z$ directly into a Taylor series instead of the sky-plane $x$ and $y$ positions. However,
the projected distance has a relatively sharp minimum (compared to the behaviour of $x$ and $y$ positions), and the
time of the minimum does not necessarily match our mid-transit time for which $x=0$. Thus, expanding $z$ would require
one to first find the minimum $z$ time and then include higher-order derivatives into the series. This increases
complexity of the implementation and would also likely reduce numerical stability, so we decided to prefer
the approach described here.}

The approach naturally works when modelling transits (or eclipses) only, and the full Keplerian orbit needs to be 
evaluated when modelling phase curves. However, even then it may be beneficial to calculate the projected distances 
for the transit model using the Taylor series approach, especially if the transit model needs to be supersampled.

\change{The planet-star contact points (beginning of ingress, $T_1$, end of ingress, $T_2$, beginning of egress, 
$T_3$, and end of egress $T_4$, \citealt{Winn2010e}) are easy to compute numerically. Calculation of a single point takes 
$\approx 500$~ns, and the calculation of different durations ($T_{14}$, $T_{23}$, $T_{12}$, and $T_{23}$) 
takes between 1-2~$\mu$s.
We do not include the code to calculate the contact points here, but make it available
from \textsc{PyTransit} repository in \textsc{GitHub}. \textsc{PyTransit} also uses the $T_1$ and $T_4$ points
to create a transit bounding box in time that is used to ensure we do not waste time evaluating the model over the 
out-of-transit points.}

The centre time for the series expansion, $t_0$, affects the accuracy. It could be beneficial to choose $t_0$ to
match the time where $y$ is minimum (so that $y$ velocity is zero), or the time of minimum $z$. It could
also be possible to choose a different $t_0$ for the $x$ and $y$ expansion. However, both approaches would
require more computation to solve those locations than just choosing the mid-transit time, and are probably
not worth the work considering that the current approach already reaches an accuracy that has basically no
effect on the transit light curve model.

The speed gains discussed in Sect.~\ref{sec:performance} depend significantly on the overall implementation of the
whole transit model. The examples in this study consider light curves where most of the points are in transit 
(that is, most of the out-of-transit data has been removed). Having a light curve with a small fraction of 
in-transit points (such as when modelling a full \textit{Kepler} or \textit{TESS} light curve directly) will
significantly increase the speed gain unless the transit model is smart enough to skip the out-of-transit 
points.

The final effect on the evaluation speed also depends on the other parts of the posterior computation, such as
the noise model. The gain will be smaller when the posterior computation time is dominated by the noise
model evaluation (such as when using brute-force Gaussian Processes), and greatest in an analysis with a
computationally cheap noise model and a large number of data points.

\change{The approach has been adopted as the main $z$ computation method in the \textsc{PyTransit} exoplanet 
transit modelling package by \citet{Parviainen2015}. However, considering the simplicity of the approach, we 
believe it can be useful to 
everyone developing exoplanet transit models and modelling frameworks independent of the programming language 
used. Thus, the approach can be easily added to other commonly used transit modelling packages, such as 
\textsc{EXOFAST} by \citet{Eastman2013}, \textsc{batman} by \citet{Kreidberg2015}, \textsc{ellc} by 
\citet{Maxted2016}, or \textsc{TLCM} by \citet{Csizmadia2020}. 
}

\section*{Acknowledgements}
We thank E. Agol for his valuable comments that substantially helped to improve the manuscript.
HP acknowledges financial support from the Agencia Estatal de Investigación del Ministerio de Ciencia, Innovación y
Universidades (MICIU) and Unión Europea Fondos FEDER (EU FEDER) funds through the project PGC2018-098153-B-C31.
JK acknowledges support by DFG grants PA525/19-1 and PA525/18-1 within the DFG Schwerpunkt SPP 1992, Exploring the Diversity of Extrasolar Planets.

\section*{Data Availability}
There are no new data associated with this article.

\bibliographystyle{mnras}
\bibliography{2020_04_taylor_z}

\appendix
\section{Example implementation}
\label{sec:implementation}

Here we show an example \texttt{numba}-accelerated \texttt{Python} implementation of the approach used 
by the \texttt{PyTransit} transit modelling package. First, a method to calculate the sky-plane x and y
derivatives
\begin{lstlisting}
from numba import njit
from numpy import (arctan2, cos, sin, sqrt, floor, 
                   mod, pi)

@njit
def ta_newton_s(t, t0, p, e, w):
    offset = arctan2(sqrt(1.0-e**2)*sin(0.5*pi-w), 
                     e + cos(0.5*pi - w))
    offset -= e*sin(offset)
    ma = mod(2*pi*(t-(t0-offset*p/2*pi))/p, 2*pi)
    ea = ma
    err = 0.05
    k = 0
    while abs(err) > 1e-8 and k < 1000:
        err = ea - e*sin(ea) - ma
        ea = ea - err/(1.0-e*cos(ea))
        k += 1
    sta = sqrt(1.0-e**2)*sin(ea)/(1.0-e*cos(ea))
    cta = (cos(ea)-e)/(1.0-e*cos(ea))
    return arctan2(sta, cta)

@njit
def xyeo(t, t0, p, e, w, ae, ci):
    f = ta_newton_s(t, t0, p, e, w) 
    r = ae / (1.+ e*cos(f))
    x = -r * cos(w + f)
    y = -r * sin(w + f) * ci
    return x, y
    
@njit
def vajs_from_paiew(t0, p, a, i, e, w):
    """Planet velocity, acceleration, jerk, and
    snap at mid-transit in [R_star / day]"""

    # Time step for central finite difference
    # ---------------------------------------
    # I've tried to choose a value that is small 
    # enough to work with USP orbits and large 
    # enough not to cause floating point problems 
    # with the fourth derivative (anything much 
    # smaller starts hitting the double precision
    # limit.)
    dt = 2e-2
    
    # Calculation of X and Y positions
    # --------------------------------
    # These could be calculated with a loop 
    # with X and Y as arrays, but I'm unrolling 
    # the loop manually because this seems to 
    # give a small speed advantage with numba.
    ae = a*(1.-e**2)
    ci = cos(i)
    
    x0, y0 = xyeo(t0-3*dt, t0, p, e, w, ae, ci)
    x1, y1 = xyeo(t0-2*dt, t0, p, e, w, ae, ci)
    x2, y2 = xyeo(t0-1*dt, t0, p, e, w, ae, ci)
    x3, y3 = xyeo(t0     , t0, p, e, w, ae, ci)
    x4, y4 = xyeo(t0+1*dt, t0, p, e, w, ae, ci)
    x5, y5 = xyeo(t0+2*dt, t0, p, e, w, ae, ci)
    x6, y6 = xyeo(t0+3*dt, t0, p, e, w, ae, ci)

   # First time derivative of position: velocity
    # -------------------------------------------
    a, b, c = 1/60, 9/60, 45/60
    vx=(a*(x6-x0)+b*(x1-x5)+c*(x4-x2))/dt
    vy=(a*(y6-y0)+b*(y1-y5)+c*(y4-y2))/dt

    # Second time derivative: acceleration
    # ------------------------------------
    a, b, c, d = 1/90, 3/20, 3/2, 49/18
    ax=(a*(x0+x6)-b*(x1+x5)+c*(x2+x4)-d*x3)/dt**2
    ay=(a*(y0+y6)-b*(y1+y5)+c*(y2+y4)-d*y3)/dt**2

    # Third time derivative: jerk
    # ---------------------------
    a, b, c = 1/8, 1, 13/8
    jx=(a*(x0-x6)+b*(x5-x1)+c*(x2-x4))/dt**3
    jy=(a*(y0-y6)+b*(y5-y1)+c*(y2-y4))/dt**3

    # Fourth time derivative: snap
    # ----------------------------
    a, b, c, d = 1/6, 2, 13/2, 28/3
    sx=(-a*(x0+x6)+b*(x1+x5)-c*(x2+x4)+d*x3)/dt**4
    sy=(-a*(y0+y6)+b*(y1+y5)-c*(y2+y4)+d*y3)/dt**4

    return y3, vx, vy, ax, ay, jx, jy, sx, sy
\end{lstlisting}
Here \texttt{xyeo} calculates the x and y positions at
given times using the Newton's method to calculate the true anomaly (\texttt{ta\_newton\_s}).
Now, the projected distance can be calculated using the derivatives as a Taylor series
\begin{lstlisting}
@njit(fastmath=True)
def z_taylor(tc, t0, p, y0, 
             vx, vy, ax, ay, 
             jx, jy, sx, sy):
    """Projected planet-star distance using a 
    Taylor series expansion."""
    
    epoch = floor((tc - t0 + 0.5*p) / p)
    t = tc - (t0 + epoch * p)
    t2 = t*t
    t3 = t*t2
    t4 = t*t3
    px =      vx*t + ax*t2/2 + jx*t3/6 + sx*t4/24
    py = y0 + vy*t + ay*t2/2 + jy*t3/6 + sy*t4/24
    return sqrt(px**2 + py**2)
\end{lstlisting}

\bsp
\label{lastpage}
\end{document}